\documentclass[twocolumn,aps,pra,reprint,superscriptaddress,longbibliography]{revtex4-1}

\usepackage{amsmath}
\usepackage{graphicx}
\usepackage[lofdepth,lotdepth, caption=false]{subfig}
\usepackage{verbatim}
\usepackage{color}
\usepackage{dcolumn}
\usepackage{bm}
\usepackage{miller}
\usepackage{color}
\usepackage{xr-hyper}
\usepackage{hyperref}
\usepackage[english]{babel}

\begin{document}

\title{Lattice and magnetic dynamics in YVO$_{3}$ Mott insulator studied by neutron scattering and first-principles calculations} 

\author{Yu Tao}
%\email{yt3ba@virginia.edu}
\affiliation{Department of Physics, University of Virginia, Charlottesville, Virginia 22904, USA}

\author{Douglas L.\ Abernathy}
\affiliation{Neutron Scattering Division, Oak Ridge National Laboratory, Oak Ridge, Tennessee 37831, USA}

\author{Tianran Chen}
\affiliation{NIST Center for Neutron Research, National Institute of Standards and Technology, Gaithersburg, Maryland 20877, USA}

\author{Taner Yildirim}
\affiliation{NIST Center for Neutron Research, National Institute of Standards and Technology, Gaithersburg, Maryland 20877, USA}
\affiliation{Materials Science and Engineering, University of Pennsylvania, Philadelphia, PA 19104, USA}

\author{Jiaqiang Yan}
\affiliation{Materials Science and Technology Division, Oak Ridge National Laboratory, Oak Ridge, Tennessee 37831, USA}
\affiliation{Department of Physics and Astronomy, University of Tennessee, Knoxville, Tennessee 37996, USA}

\author{Jianshi Zhou}
\affiliation{Department of Mechanical Engineering, The University of Texas at Austin, Austin, TX, USA}

\author{John B. Goodenough}
\affiliation{Department of Mechanical Engineering, The University of Texas at Austin, Austin, TX, USA}

%\author{Despina Louca$^{\bullet}$(corresponding author)}
\author{Despina Louca}
\thanks{Corresponding author}
\email{louca@virginia.edu}
\affiliation{Department of Physics, University of Virginia, Charlottesville,
Virginia 22904, USA}

\begin{abstract}

The Mott insulator YVO$_{3}$ with $T_{N}$ = 118 K is revisited to explore the role of spin, lattice and orbital correlations across the multiple structural and magnetic transitions observed as a function of temperature. Upon cooling, the crystal structure changes from orthorhombic to monoclinic at 200 K, and back to orthorhombic at 77 K, followed by magnetic transitions. From the paramagnetic high temperature phase, C-type ordering is first observed at 118 K, followed by a G-type spin re-orientation transition at 77 K. The dynamics of the transitions were investigated via inelastic neutron scattering and first principles calculations. An overall good agreement between the neutron data and calculated spectra was observed. From the magnon density of states, the magnetic exchange constants were deduced to be $J_{ab}$ = $J_{c}$ = -5.8 meV in the G-type spin phase, and $J_{ab}$ = -3.8 meV, $J_{c}$ = 7.6 meV at 80 K and $J_{ab}$ = -3.0 meV, $J_{c}$ = 6.0 meV at 100 K in the C-type spin phase. Paramagnetic scattering was observed in the spin ordered phases, well below the C-type transition temperature, that continuously increased above the transition. Fluctuations in the temperature dependence of the phonon density of states were observed between 50 and 80 K as well, coinciding with the G-type to C-type transition. These fluctuations are attributed to optical oxygen modes above 40 meV, from first principles calculations. In contrast, little change in the phonon spectra is observed across $T_{N}$.

\end{abstract}
\maketitle

\section{Introduction}

Transition metal oxides with the perovskite structure exhibit many interesting properties due to strong electron correlations leading to a Mott transition \cite{Cyrot1972, Shen1995}, high-temperature superconductivity \cite{Shen1995}, colossal magnetoresistivity \cite{Tokura2000} and ferroelectricity \cite{Efremov2004}. Included in this class is the insulating YVO$_{3}$ in which correlation effects can lead to Jahn-Teller (JT) distortions, orbital ordering, charge and spin stripe formation, polaron localization, and phase separation. YVO$_{3}$ has a complex phase diagram associated with its orbital physics, and it has been extensively studied \cite{Noguchi2000, Blake2001, Ulrich2003}. However, little is known of the magnetic and lattice dynamics across the many phase transitions observed in this system, which is the focus of the present work.

\begin{figure}[h]
\begin{center}
\includegraphics[width=8.6cm]
{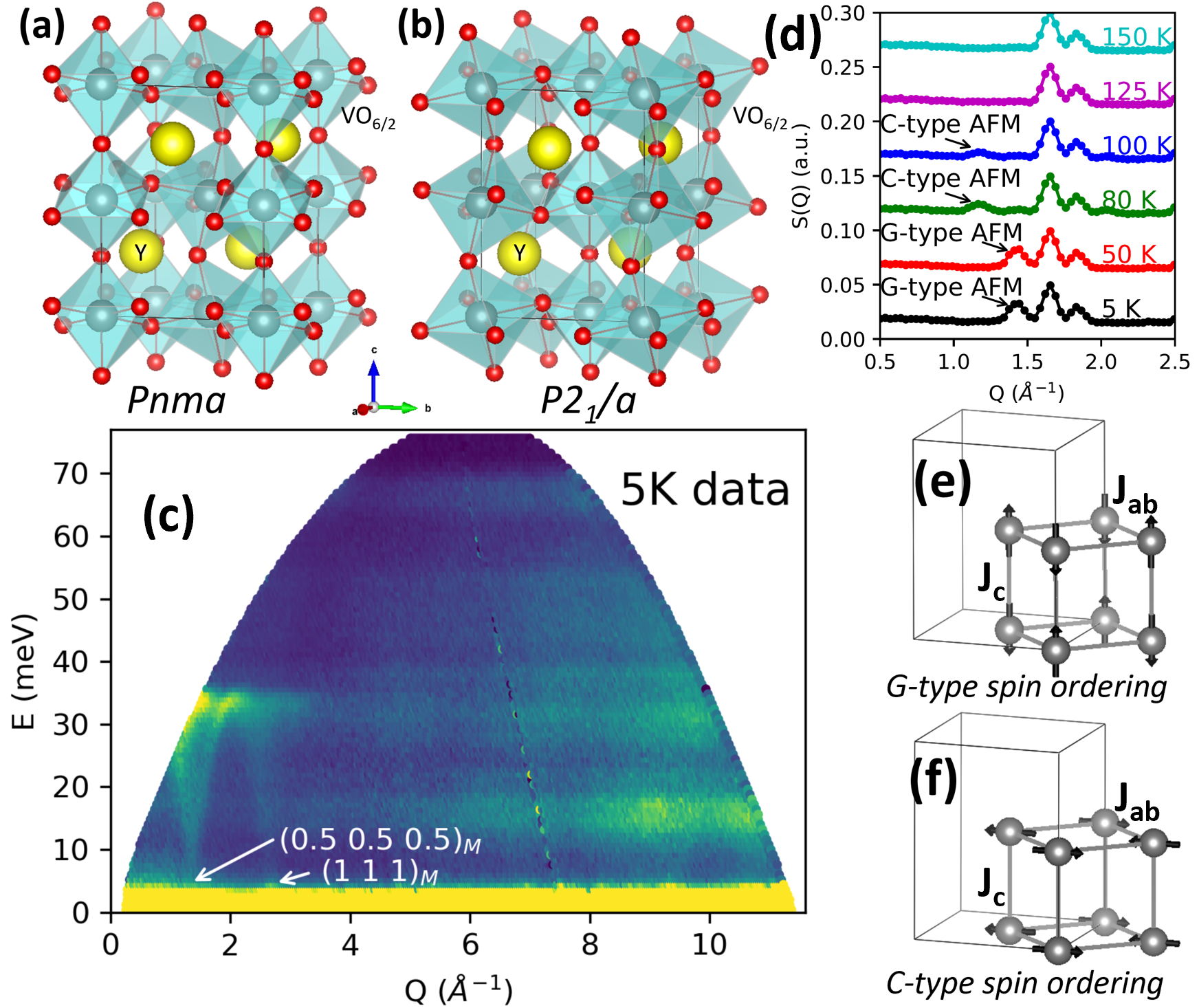}
\end{center}
\caption{The crystal structures of YVO$_{3}$ in (a) the orthorhombic phase ($Pnma$) and (b) the monoclinic phase ($P2_{1}/a$). (c) A plot of the inelastic neutron scattering intensity as a function of momentum transfer ($Q$) and energy ($E$) on a powder sample of YVO$_{3}$ at $T$ = 5 K, with an incident neutron energy of $E_{i}$ = 80 meV. (d) Elastic neutron scattering intensity versus momentum transfer as a function of temperature. The labeled AFM peaks show the magnetic transition from the G-type to the C-type spin ordered state. (e) G-type spin ordered structure in the orthorhombic unit cell (solid black line). (f) C-type spin ordered structure in the monoclinic unit cell.}
\label{fig:fig1}
\end{figure}

YVO$_{3}$ adopts an orthorhombically distorted perovskite structure with space group $Pnma$ at room temperature (Fig.\ \ref{fig:fig1}(a)). The $3d$ magnetic ions V$^{3+}$ with spin $S=1$ are doubly degenerate in the $t^{2}_{2g}$ manifold. On cooling below $T_{OO}$ = 200 K, YVO$_{3}$ first undergoes a structural phase transition from the orthorhombic to a monoclinic phase ((Fig.\ \ref{fig:fig1}(b))) with its symmetry lowered down to $P2_{1}/a$. At this temperature, the 3d orbitals of V$^{3+}$ form a G-type orbital ordering pattern along the c-axis. A second phase transition occurs at $T_{N}$ = 118 K, where the spins form C-type antiferromagnetic (AFM) structure in the ab-plane with ferromagnetic coupling along the $c$-axis (C-type spin ordering (C-SO), Fig.\ \ref{fig:fig1}(f)). Finally, with further cooling, the system transforms back to its original orthorhombic structure (with the $Pnma$ crystal symmetry) at $T_{CG}$ = 77 K. This phase transition is accompanied by a change of the spin configuration into a G-type AFM order (G-SO, Fig.\ \ref{fig:fig1}(e)), with a simultaneous change in the orbital order from G- to C-type \cite{Blake2002, Blake2001}.

In both the G- and C-type orbitally ordered phases, the $d_{xy}$ orbital is always occupied due to its lower energy, and the second occupied orbital alternates between $d_{yz}$ or $d_{xz}$ along the $c$-axis \cite{Yano2014}. Superexchange interactions between the neighboring spins and orbitals as proposed in Refs. \cite{Kugel1973, Ishihara2004} give rise to the orbital and spin ordering patterns observed in YVO$_{3}$. Experimentally, these transitions have been widely studied using several techniques including optical reflectivity \cite{Miyasaka2002}, synchrotron X-ray diffraction \cite{Noguchi2000}, specific heat \cite{Blake2002} and neutron scattering \cite{Reehuis2006, Ren1998, Blake2001}.

Previous neutron diffraction measurements combined with the analysis of the local structure using the pair distribution function (PDF) technique provided evidence for a local G-type orbital ordering pattern, where the local structure favors a modified $P2_{1}/a$ symmetry. The presence of a local monoclinic pattern in the overall orthorhombic structure suggests that the transitions in YVO$_{3}$ are more complex and worth further investigation \cite{Yano2014}. Although the static atomic and magnetic structures of YVO$_{3}$ have been well characterized, there are not many studies that address the dynamics of the transitions and the nature of the magneto-orbital mechanism. Coupled with the structural transitions are steric VO$_{3}$ octahedral rotations that have direct consequences on the magnetic and phonon dynamics. The dynamics of the rotations have not been probed, to the best of our knowledge. Therefore, investigating the phonon spectra at different temperatures across the transition will elucidate the nature of the phase transitions observed in YVO$_{3}$. Although Raman spectroscopy has been used to probe the phonon and magnon changes in YVO$_{3}$, the technique provides only zone center information \cite{Sugai2006, Miyasaka2006, Jandl2010}. Extending this beyond the Brillouin zone center requires techniques such as inelastic neutron scattering. Of the sixty phonon modes, only twenty four of them are Raman active. Neutron scattering will complement much needed information on the phonons and spin spectra in YVO$_{3}$.

In this paper, we report on the spin-wave and phonon dynamics determined from powder inelastic neutron scattering measurements and first-principles calculations. The transition from the G- to C-type spin ordered states is accompanied by clear changes in the magnon spectra on warming between 50 and 80 K. The magnetic exchange constants in the G- and C-type spin ordered structures were extracted to be $J_{ab}$ = $J_{c}$ = -5.8 meV at 5 and 50 K, $J_{ab}$ = -3.8 meV, $J_{c}$ = 7.6 meV at 80 K, and $J_{ab}$ = -3.0 meV, $J_{c}$ = 6.0 meV at 100 K obtained from spin-wave analysis. Below the magnetic transition of 118 K, paramagnetic scattering is present in the dynamic structure function S($Q$,$E$). The phonon spectra up to 130 meV as a function of temperature was also extracted from the data and compared to the ones calculated from first-principles. The phonon density of states changes between 50 and 80 K, and these were attributed to energy fluctuations of optical oxygen modes above 40 meV, linked to the orthorhombic to monoclinic structural transition on warming at 77 K. By comparison, little change in the phonon spectra was observed across $T_{N}$, from C-type spin ordering, to the high temperature paramagnetic phase.

\section{Experimental and Calculation Details}

A 30 g powder sample of YVO$_{3}$ was prepared for the measurement. Single-phase crystals were melt-grown using an image furnace following the procedure described elsewhere \cite{Yan2004}. The powder sample was obtained by crushing the single crystals. The inelastic neutron scattering experiment was performed on the wide angular-range chopper spectrometer ARCS, located at the Spallation Neutron Source of Oak Ridge National Laboratory (ORNL). The measurements used incident neutron energies of 80 and 130 meV. The powder was loaded into an aluminum can. The data were collected as a function of temperature between 5 and 150 K. In this work, we use atomic coordinates with $a < b < c$, i.e., $a \approx$ 5.287 \AA, $b \approx$ 5.594 \AA, and $c \approx$ 7.560 \AA \ for the orthorhombic unit cell, $a \approx$ 5.274 \AA, $b \approx$ 5.605 \AA, $c \approx$ 7.544 \AA \ and $\alpha \approx$ 90.032$^{\circ}$ for the monoclinic unit cell.

The density functional theory (DFT) calculations were performed using the VASP Package \cite{hafner2008ab} implementing local-density approximations (LDA) exchange-correlation functional. The simplified (rotationally invariant) approach to the LSDA+U introduced by Dudarev et al. \cite{dudarev1998electron} was implemented to correct the approximate exchange-correlation functional. The factor U which specifies the strength of the effective on-site Coulomb interactions was set to 4 eV.

The k-space integration was performed with 8$\times$8$\times$8 k points in the Brillouin zone for the structural optimization. For the structural relaxation in the G-type phase, atoms were allowed to relax along the calculated forces until all the residual force components were less than 1e$^{-7}$ eV/a.u. In the C-type phase, to make sure the phonon changes only come from the spin configuration difference but not from the structural difference, the crystal structure was kept the same as in G-type and only the spin configurations were changed. The phonon band structures and density of states were calculated using the finite difference method implemented by the Phonopy software \cite{togo2015first} and VASP Package using a 2$\times$2$\times$2 supercell and 4$\times$4$\times$3 k-grids.

\section{Results and Discussion}

The time-of-flight data were collected as a function of temperature from 5 to 150 K. Shown in Fig.\ \ref{fig:fig1}(c) is the $E-Q$ contour map of the dynamic structure function, S($Q$,$E$), at 5 K using an incident energy $E_{i}$ = 80 meV. The data were reduced, analyzed and visualized using DAVE \cite{Azuah2009}. The background and empty can were also subtracted. With this measurement both magnon and phonon excitations are captured. At low Q, magnetic dispersions are visible. At high Q, the horizontal bright bands, with the intensity increasing as a function of $Q^{2}$ are due to the phonon integrated dispersion spectrum. In the low-$Q$ region, magnon dispersions are clearly visible up to 4 \AA$^{-1}$, corresponding to the G-type spin ordering below 77 K. The magnetic form factor suppresses the intensity beyond 4 \AA$^{-1}$.

The integrated elastic scattering intensity between $Q$ = 0.5 and 2.5 \AA$^{-1}$ is plotted in Fig.\ \ref{fig:fig1}(d) as a function of temperature. The cuts were obtained by integrating from -5 to 5 meV across the elastic line. Indicated on this figure are the positions in $Q$ of the static magnetic Bragg peaks. From Fig.\ \ref{fig:fig1}(c), it can be seen that the spin-wave excitations emanate from the Bragg peaks marked in Fig.\ \ref{fig:fig1}(d). For the elastic data shown at 5 and 50 K, a Bragg peak appears at $\approx$ 1.4 \AA$^{-1}$, corresponding to the (0.5, 0.5, 0.5)$_{M}$ G-type AFM structure. Increasing the temperature to 80 K, a C-type magnetic AFM peak at (0.5, 0.5, 0)$_{M}$ appears at $\approx$ 1.2 \AA$^{-1}$, that follows the G- to C-type spin ordering transition at 77 K. This can be seen in the contour map of Fig.\ \ref{fig:fig3}(c). The magnetic peaks disappear above $T_{N}$.

\subsection{Spin Wave Analysis}

\begin{figure}[h]
\begin{center}
\includegraphics[width=8.6cm]
{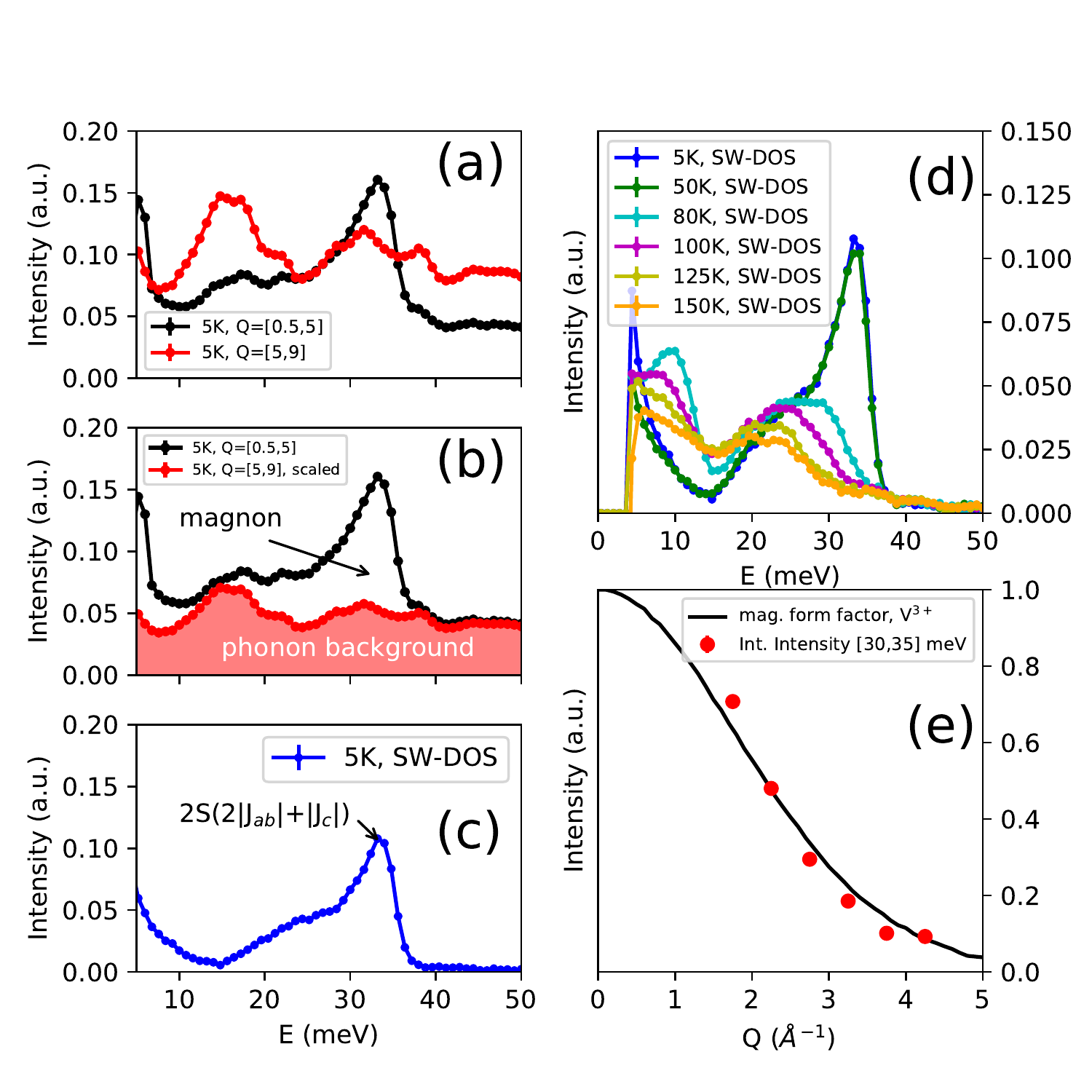}
\end{center}
\caption{(a) Inelastic neutron scattering intensity integrated over the low-$Q$ region (LQ) from 0.5 to 5 \AA$^{-1}$, and over the high-$Q$ region (HQ) from 5 to 9 \AA$^{-1}$. (b) The phonon background, determined from the HQ cut, is subtracted from the LQ data. (c) The intensity difference between LQ and HQ cuts is the magnon DOS at 5 K. (d) The extracted magnon DOS versus energy transfer as a function of temperature between 5 and 150 K. (e) The integrated intensity between 30 and 35 meV at 5 K as a function of $Q$. The results are compared with the calculated magnetic form factor of V$^{3+}$.}
\label{fig:fig2}
\end{figure}

The dynamic structure function S($Q$,$E$) as a function of momentum transfer ($Q$) and energy ($E$) consists of both magnon and phonon information. To separate the two contributions, data were integrated in $Q$ from 0.5 to 5 \AA$^{-1}$ (the low-$Q$ cut (LQ)) and from 5 to 9 \AA$^{-1}$ (the high-$Q$ cut (HQ)).The two cuts are shown in Fig.\ \ref{fig:fig2}(a). The LQ-cut includes the entire magnon spectrum based on the magnetic form factor calculation for V$^{3+}$ (Fig.\ \ref{fig:fig2}(e)), plus low energy phonons. The HQ-cut, on the other hand, should be devoid of spin-waves, and be mostly due to phonons. The method used to remove the phonon contribution was the same as the one performed in Ref.\ \cite{McQueeney2008}, for LaMnO$_{3}$, LaVO$_{3}$ and LaFeO$_{3}$. The in-plane $J_{ab}$ and out-of-plane $J_{c}$ can be determined from the positions of the magnon peaks. This method has also been used in several other magnetic systems \cite{Delaire2012, Mishra2016, Johnston2011, Ramazanoglu2017}.

To subtract the phonon background, the HQ-cut was scaled by a factor, in this case, k = 0.48, so that the high energy region of the HQ-cut overlaps with the high energy region of the LQ-cut. This is because phonon intensity is proportional to $Q^{2}$ and this is clearly seen in the HQ-cut. In the high energy region, the LQ-cut only contains low energy phonons. This phonon background is then subtracted from all the data at each temperature, as shown in Fig.\ \ref{fig:fig2}(b), and the remaining intensity is due to magnons as shown in Fig.\ \ref{fig:fig2}(c).

Shown in Fig.\ \ref{fig:fig2}(d) is a plot of the spin-wave density of states (SW-DOS) as a function of temperature. The elastic intensity at each temperature was fitted with a Gaussian peak centered at 0 meV and subtracted from the data. The magnetic structure of YVO$_{3}$ is described in terms of a pseudocubic cell with lattice parameters of $a$/$\sqrt2$, $b$/$\sqrt2$ and $c$/2 \cite{Ulrich2003}. The magnetic exchange coupling constants between neighboring spins are extracted from the SW-DOS.

In the SW-DOS, the magnon peaks indicate the energy extrema corresponding to van Hove singularities in the dispersion curves that are related to the magnetic exchange coupling constants $J_{ab}$ and $J_{c}$ \cite{McQueeney2008}. For example, the magnon peak at $\approx$ 35 meV in the 5 K SW-DOS has an energy equivalent to $2S(2|J_{ab}|+|J_{c}|)$ (Fig.\ \ref{fig:fig2}(c)). In the G-type spin ordered state, little change is observed in the SW-DOS between 5 and 50 K. The anisotropic magnetic exchange coupling constants were determined to be $J_{ab}$ = $J_{c}$ = -5.8 meV at both temperatures. These values are consistent with the reported values of $J_{ab}$ = $J_{c}$ = -5.7 meV, measured using single crystal neutron scattering \cite{Ulrich2003}.

In the C-type spin ordering state, two magnon peaks are expected. The first peak is located at $E$ $\approx$ 10 meV, which corresponds to $4S|J_{ab}|$. The second peak is located at $E$ $\approx$ 30 meV, which corresponds to $4S(|J_{ab}|+|J_{c}|)$. When $T$ $>$ 77 K, a large reduction of the ordered moment of V$^{3+}$ due to strong quantum fluctuations is expected. The effective moment of V$^{3+}$ from the refinement of the magnetic Bragg peaks is between 1.05 and 1.11 $\mu_{B}$, much lower than the free-ion moment of 2 $\mu_{B}$ \cite{Blake2002, Ulrich2003, Sawada1998}. From the SW-DOS data at 80 K, the magnetic exchange coupling constants using the refined effective moment were extracted to be $J_{ab}$ = -3.8 meV, $J_{c}$ = 7.6 meV. At 100 K, both magnon peaks are damped, in contrast to the G-type spin ordering state. The magnon peaks at 100 K are located at $E$ $\approx$ 8 and 24 meV, indicating a decrease of $J_{ab}$ to -3.0 meV and $J_{c}$ to 6.0 meV. At 125 and 150 K in the paramagnetic (PM) state, the magnon peaks are further damped, but still present, indicating that magnetic correlations still survive in the PM phase.

In Fig.\ \ref{fig:fig2}(e), the change of the integrated magnon intensity at 5 K is plotted as a function of $Q$. The integration was performed between 30 and 35 meV, with a $Q$ width of 0.5 \AA$^{-1}$. The magnetic form factor of V$^{3+}$ was computed and is also plotted for comparison. It is clear that the magnon intensity follows the $Q$-dependence of the magnetic form factor.

\begin{figure}[h]
\begin{center}
\includegraphics[width=8.6cm]
{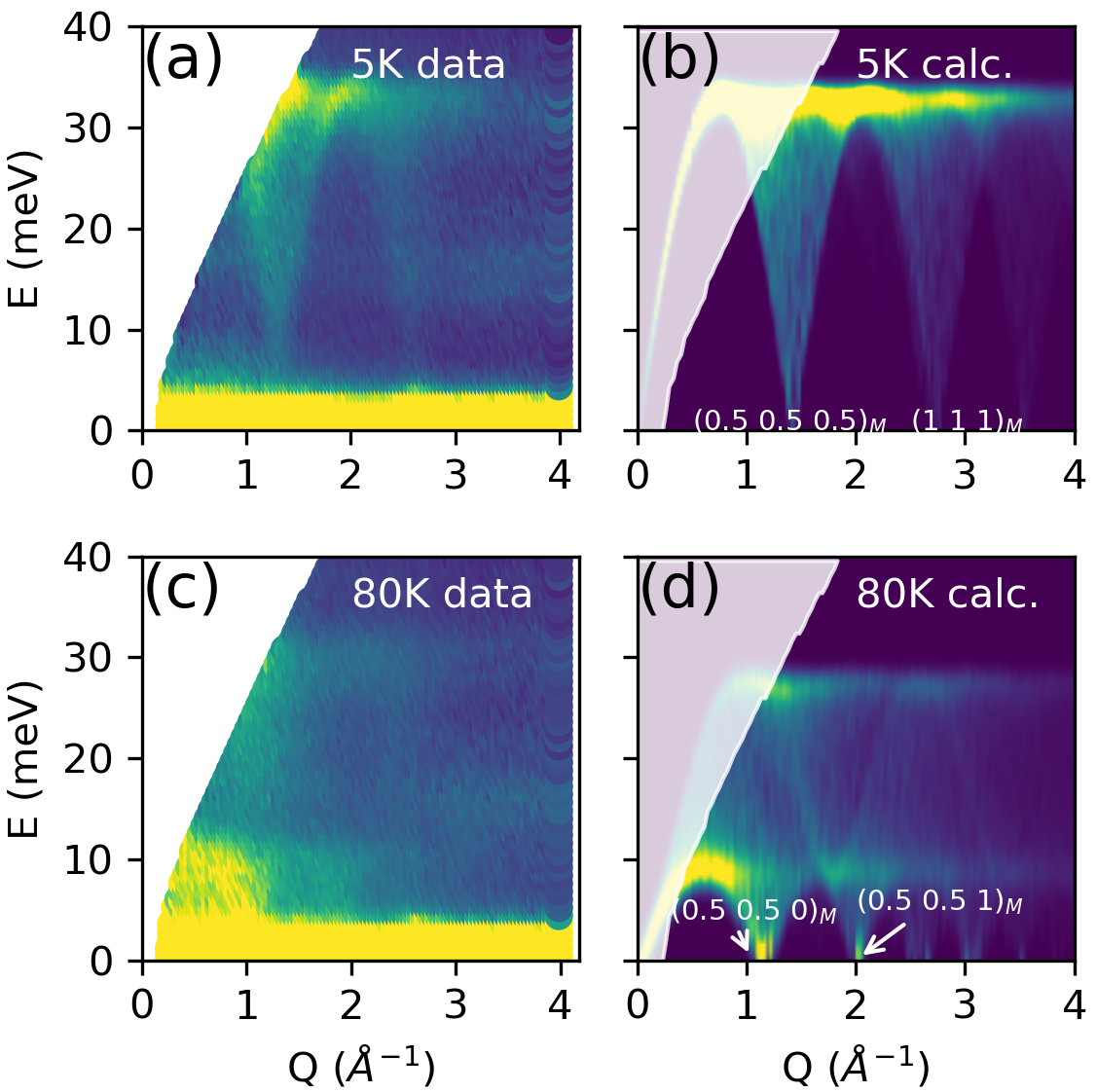}
\end{center}
\caption{(a, c) Inelastic neutron scattering intensity S($Q$,$E$) of YVO$_{3}$ at 5 and 80 K. The data panels are shown over a smaller $Q$ and $E$ range as compared to Fig.\ \ref{fig:fig1}(e), and include mainly the spin waves. (b, d) Calculated powder averaged spin wave dispersions of YVO$_{3}$ using a Heisenberg AFM model with G- or C-type spin ordering at 5 and 80 K, convoluted with the resolution function of the ARCS instrument.}
\label{fig:fig3}
\end{figure}

At 5 K, the G-type spin-waves emanate from the AFM Bragg peaks of (0.5 0.5 0.5)$_{M}$ and (1 1 1)$_{M}$, located at $Q$ $\approx$ 1.4 and 2.8 \AA$^{-1}$. The maximum magnon energy is at $\approx$ 35 meV (Fig.\ \ref{fig:fig3}(a)). At 80 K, C-type magnetic Bragg peaks were observed at (0.5 0.5 0)$_{M}$ at Q $\approx$ 1.2 \AA$^{-1}$ and (0.5 0.5 1)$_{M}$ at $Q$ $\approx$ 2.0 \AA$^{-1}$. A lower magnon band is at $\approx$ 10 meV along with a higher magnon band at $\approx$ 30 meV (Fig.\ \ref{fig:fig3}(c)).

The simulated spin-waves at 5 K G-type and 80 K C-type spin ordered states are shown in Fig.\ \ref{fig:fig3}(b,d), and are compared with the neutron data in panels (a,c). The calculations were performed using the SpinW software \cite{Toth2015}. The Hamiltonian of a simple Heisenberg AFM model ($-J_{ab} \sum_{<i,j>||a,b} S_{i} \cdot S_{j}-J_{c} \sum_{<i,j>||c} S_{i} \cdot S_{j}$) was used for both magnetic structures. Reported lattice parameters from earlier studies were used \cite{Yano2014}. The calculated magnon intensity was further convoluted with the instrument resolution function. The simulated patterns reproduce the spin-wave spectra well.

Even for temperatures below $T_{N}$, PM scattering is still observed in both spin ordered phases. From the SW-DOS data at 5 and 50 K shown in Fig.\ \ref{fig:fig2}(d), the PM intensity is observed between 5 and 15 meV. At 80 and 100 K, some inelastic intensity appears in the low-Q region between 0 and 1 \AA$^{-1}$, and from 0 to 10 meV, as shown in Fig.\ \ref{fig:fig3}(c). This PM intensity is not reproduced by the simulation, and suggests that short-range PM spin correlations might persist in YVO$_{3}$ despite the long-range magnetic order.

Magnon-magnon interactions might also exist in YVO$_{3}$. As shown in Figs.\ \ref{fig:fig3}(a,c), the spin-waves at 80 K are less sharply defined compared with the data at 5 K. Upon warming, the C-type magnon peaks in Fig.\ \ref{fig:fig2}(d) gradually shift to lower energy and become strongly damped just below the magnetic transition. In contrast, the magnons in the G-type spin ordered state show little temperature dependence. Evidence of magnon-magnon interactions in YVO$_{3}$ has also been provided by Raman spectroscopy \cite{Sugai2006}. Above $T_{N}$, the spin-waves are replaced by a broad PM scattering intensity. The magnon-magnon interactions, together with the PM scattering suggest that the C-type spin ordered phase is much more complex than the G-type.

\begin{figure}[h]
\begin{center}
\includegraphics[width=8.6cm]
{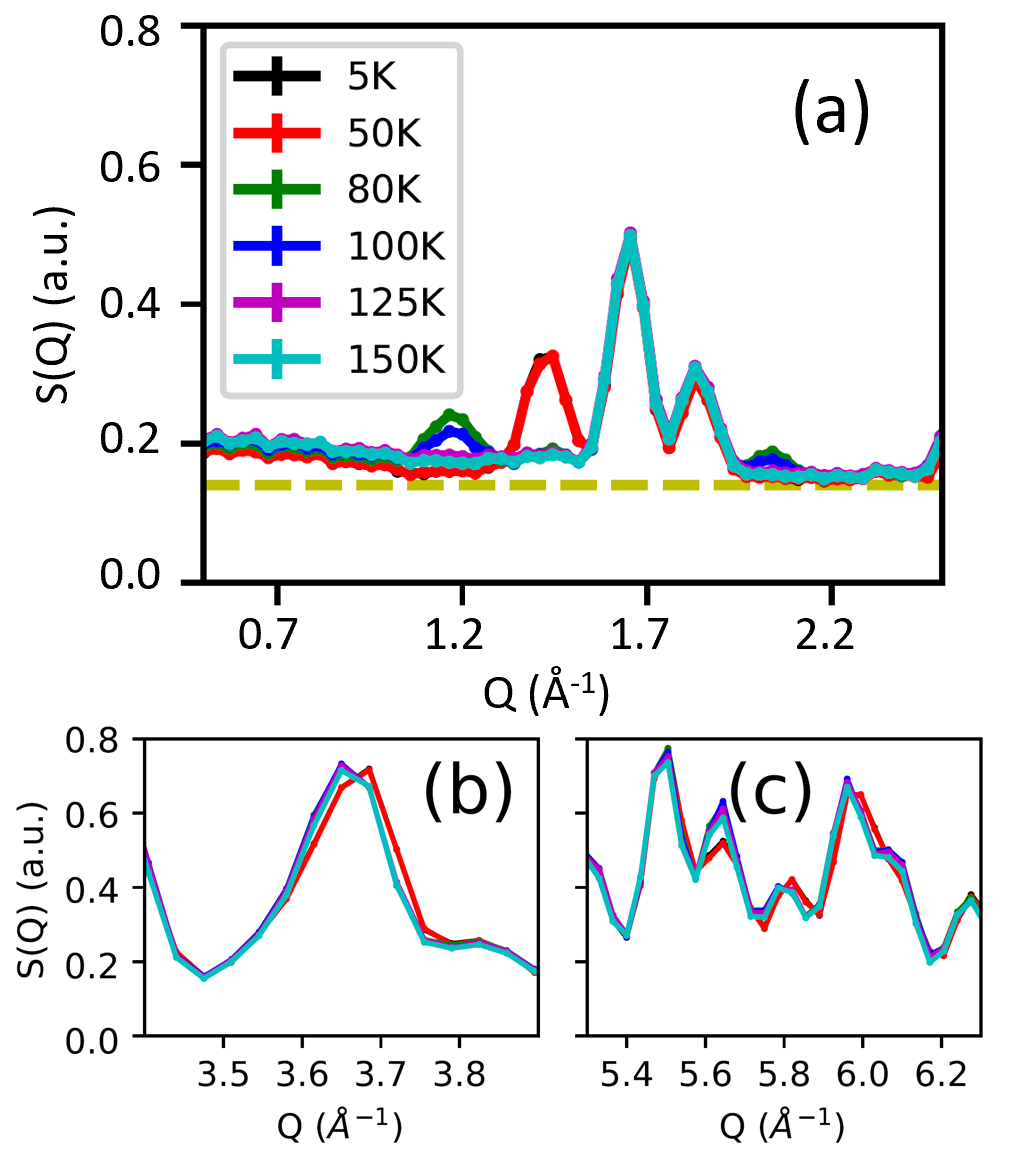}
\end{center}
\caption{Elastic neutron scattering intensity versus momentum transfer as a function of temperature at different Q ranges: (a) 0.5 \AA$^{-1}$ $\leq$ $Q$ $\leq$ 2.5 \AA$^{-1}$, (b) 3.4 \AA$^{-1}$ $\leq$ $Q$ $\leq$ 3.9 \AA$^{-1}$, (c) 5.3 \AA$^{-1}$ $\leq$ $Q$ $\leq$ 6.3 \AA$^{-1}$. The dashed line in panel (a) indicate the background level, with clear PM scattering intensity present below 1.5 \AA$^{-1}$.}
\label{fig:fig4}
\end{figure}

While most studies carried out so far indicate a single magnetic phase (C-type) between 77 and 118 K \cite{Reehuis2006, Blake2001, Ulrich2003}, two recent studies using neutron diffraction and synchrotron X-ray diffraction provided evidence of phase coexistence (C- and G-type) in powder samples in this temperature range and a second order nature of the magnetic transition at $T_{N}$ \cite{Sharma2021, Saha2017}. The phase fraction of the G-type spin ordered state decreases gradually with temperature, from 96\% at 80 K to 50\% at 100 K, replaced by the C-type spin ordered state, with clear changes of the Bragg peak intensity \cite{Sharma2021}. However, such coexistence is not observed in our data. In Fig.\ \ref{fig:fig4}, the elastic neutron scattering intensity as a function of temperature is plotted in three different Q ranges. Changes of the Bragg peak intensity were only observed between 50 and 80 K, corresponding to the orthorhombic to monoclinic transition at 77 K. Little change was observed between 80 and 100 K, from both the nuclear and the magnetic Bragg peaks. In Fig.\ \ref{fig:fig4}(a), no G-type magnetic Bragg peaks were present at 80 or 100 K, and the higher background below 1.5 \AA$^{-1}$ indicates that PM scattering is present together with a long-range C-type magnetic ordering. There is also no sign of the G-type spin-wave in the 80 K magnon data as shown in Fig.\ \ref{fig:fig3}(c). The reported phase coexistence might be related to sample-dependent factors. The powder used in our experiment was crushed from single crystals, while the powder used in Ref.\ \cite{Sharma2021, Saha2017} was obtained from hydrogen reduction method. The presence of grain boundaries and interparticle strain may frustrate each grain's G- to C-type magnetic transition. Different sample-dependent transition behaviors were also observed in many other materials such as WTe$_{2}$ and CrCl$_{3}$, where broader transitions were observed in powder samples compared to single crystals \cite{Tao2020, Schneeloch2021}.

\subsection{Phonon Analysis}

\begin{figure}[h!]
\begin{center}
\includegraphics[width=8.5cm]
{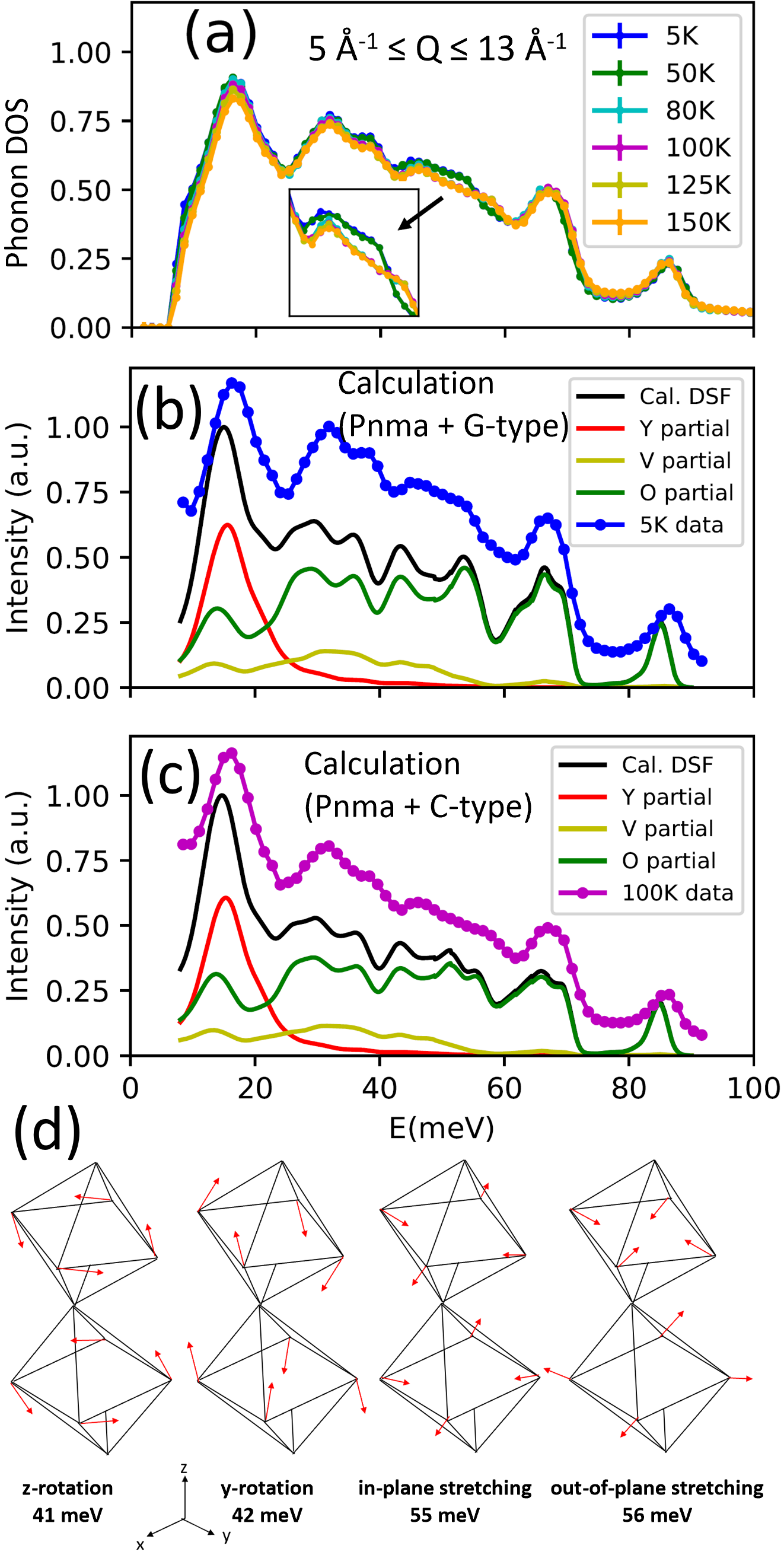}
\end{center}
\caption{(a) Generalized YVO$_{3}$ phonon DOS measured at ARCS, averaged over a $Q$ range of 5 to 13 \AA$^{-1}$, at 5, 50, 80, 100, 125 and 150 K. Inset: phonon DOS zoomed-in between 40 and 60 meV. (b) First-principles calculations of the phonon dynamical structural factor (DSF) of $Pnma$ YVO$_{3}$ at 5 K, using the G-type magnetic structure, and the partial DSFs from the Y, V, O atoms. (c) First-principles calculations of the phonon DSF of $Pnma$ YVO$_{3}$ at 100 K, with the magnetic structure changed to C-type. (d) Visualizations of some phonon modes at the $\Gamma$ point between 40 and 60 meV: $z$-rotation mode at $\approx$ 41 meV, $y$-rotation mode at $\approx$ 42 meV, in-phase stretching mode at $\approx$ 55 meV, out-of-phase stretching mode at $\approx$ 56 meV. Red arrows indicate the vibrating direction of the oxygen atoms.}
\label{fig:fig5}
\end{figure}

The S($Q$,$E$) data collected with an incident energy of 130 meV were analyzed to extract the phonon DOS as shown in Fig.\ \ref{fig:fig5}(a). The data were integrated from 5 \AA$^{-1}$ $\leq$ $Q$ $\leq$ 13 \AA$^{-1}$ in order to minimize the magnetic contribution. The measurement of the empty can was subtracted from the data and Bose corrections were performed on the data at all temperatures. The elastic peak fitted with a Gaussian function was subtracted from the data as well.

The phonon DOS changes as function of temperature. Below 40 meV, the intensity of the phonon peaks decreases gradually upon warming while the energy remains fixed. Between 40 and 60 meV, some phonon modes shift to higher energies, resulting in an overall change of the shape of the DOS. Above 60 meV, two phonon peaks show opposite temperature dependence. The peak at $\approx$ 65 meV is slightly hardened at 80 K compared with 50 K, whereas the peak at $\approx$ 85 meV undergoes a small softening from 50 to 80 K. Upon warming from 100 to 125 K, little changes in phonon spectra were observed, despite the magnetic transition at $T_{N}$ = 118 K.

To investigate the active phonon modes that are involved in the structural and spin ordered transitions at $T_{CG}$ = 77 K, density-functional theory calculations were performed to obtain the theoretical phonon DOS. In order to have a better comparison with experiment, we calculated the generalized dynamical structure factor (DSF) with the same $Q$-range averaging and energy-resolution as the experiment, and the results are shown in Fig.\ \ref{fig:fig5}(b,c) for G-type and C-type spin ordering, respectively. In order to better understand any spin-phonon coupling in YVO$_{3}$, the crystal structure is fixed to the orthorhombic $Pnma$ symmetry with the same lattice constants for both spin configurations. The partial DSFs are also shown for each type of atoms. The calculations are compared with data at 5 and 100 K (without Bose correction).

The calculated phonon DSF agrees well with the data at 5 K and allows the identification of the phonon modes that are associated with the changes observed between 50 and 80 K. Based on the partial DSFs, the broad intensity between 40 and 60 meV mainly comes from the optical oxygen modes, while little change is observed in both the Y and the V modes. Moreover, even though the crystal structure was kept the same for both calculations, the changes seen in the calculated DSF at $\approx$ 55 meV where only the spin configuration changes from G- to C-type also agree with the energy shifts seen in the data, which indicates that oxygen phonon modes between 40 and 60 meV might be coupled with the spins of V$^{3+}$.

While a change in the spin configuration partially explains the change in phonon DOS observed between 40 and 60 meV, the orthorhombic to monoclinic structural phase transition at 77 K might also contribute through the oxygen rotation modes shown in Fig. 5(d). From the reported coordinates, the orthorhombic and monoclinic structures of YVO$_{3}$ mainly differ in the bond lengths in the $ab$-plane and the $y$- and $z$-orientation of the VO$_{3}$ octahedra \cite{Blake2002}. Therefore, the $z$-rotation mode at $\approx$ 41 meV, the $y$-rotation mode at $\approx$ 42 meV, and the two in-plane stretching modes at $\approx$ 55 and 56 meV might be more active across the 77 K phase boundary. These energies correspond to the energies of the phonon modes at the $\Gamma$ point. Fig.\ \ref{fig:fig5}(d) shows the visualizations of these phonon modes at the $\Gamma$ point, obtained from the calculated phonon band structure at 5 K.

In addition, although the magnetic transition at $T_{N}$ from the C-type spin ordered phase to the high temperature PM phase is not accompanied by any structural transition, little change is observed in the phonon DOS between 100 and 125 K. This could indicate that the spin-phonon interactions in YVO$_{3}$ is weak at T$_{N}$.

\section{Conclusion}

The temperature dependence of the spin-waves and phonons in YVO$_{3}$ were investigated using powder inelastic neutron scattering measurements and first-principles lattice dynamics calculations. The G- to C-type spin ordered transition on warming was characterized from the magnon intensities, and the magnetic exchange constants in the G-type spin ordered ($J_{ab}$ = $J_{c}$ = -5.8 meV at 5 and 50 K), and C-type spin ordered ($J_{ab}$ = -3.8 meV, $J_{c}$ = 7.6 meV at 80 K, $J_{ab}$ = -3.0 meV, $J_{c}$ = 6.0 meV at 100 K) states were determined from the magnon peak positions. The calculated dispersion from SpinW using the Heisenberg AFM coupling provide a representation of the measured spin-waves. Below the magnetic transition temperature of $T_{N}$ = 118 K, PM scattering intensities are also observed. We additionally obtained the phonon spectra as a function of temperature, and compared with the results from first-principles calculations. The calculated DSFs agree well with the measured phonon intensity at 5 and 100 K. The phonon intensity shows a strong temperature dependence across the 77 K transition temperature, manifested as changes in some optical oxygen branches. In contrast, little phonon changes were seen between 100 and 125 K, despite across a magnetic transition at $T_{N}$ = 118 K, indicating rather small spin-phonon coupling in this system.

\section*{Acknowledgements}

This work has been supported by the Department of Energy, Grant number DE-FG02-01ER4592. This work was also partly supported by the Materials Research Science and Engineering Centers, National Science Foundation, Grant number DMR-1720595, by providing sample used in this work and by the National Institute of Standards and Technology, US Department of Commerce, in providing computing resources for DFT calculations used in this work. A portion of this research used resources at the Spallation Neutron Source, a DOE Office of Science User Facility operated by Oak Ridge National Laboratory.

\bibliography{YVO3}

\end{document}